\documentclass[%
 reprint,
 amsmath,amssymb,
 aps,
]{revtex4-1}

\usepackage{graphicx}
\usepackage{dcolumn}
\usepackage{bm}
\usepackage{ulem}
\usepackage{color}

\begin{document}

\preprint{APS/123-QED}

\title{Anatomy of nanomagnetic switching at a 3D Topological Insulator PN junction}
\author{Hamed Vakili}
\thanks{Authors contributed equally, correspondence email: hv8rf@virginia.edu}

\author{Yunkun Xie}%
\thanks{Authors contributed equally, correspondence email: hv8rf@virginia.edu}

\author{Samiran Ganguly}%

\author{Avik W. Ghosh}%
\affiliation{
Department of Physics,University of Virginia, Charlottesville, VA 22903 USA
}
\affiliation{School of Electrical and Computer Engineering,
University of Virginia, Charlottesville, VA 22903 USA}

\date{\today}

\begin{abstract}
A P-N junction engineered within a Dirac cone system acts as a gate tunable angular filter based on Klein tunneling. For a 3D topological insulator with a substantial bandgap, such a filter can produce a charge-to-spin conversion due to the dual effects of spin-momentum locking and momentum filtering. We analyze how spins filtered at an in-plane topological insulator PN junction (TIPNJ) interact with a nanomagnet, and argue that the intrinsic charge-to-spin conversion does not translate to an external gain if the nanomagnet also acts as the source contact. Regardless of the nanomagnet's position, the spin torque generated on the TIPNJ is limited by its surface current density, which in turn is limited by the bulk bandgap. Using quantum kinetic models, we calculated the spatially varying spin potential and quantified the localization of the current versus the applied bias. Additionally, with the magnetodynamic simulation of a soft magnet, we show that the PN junction can offer a critical gate tunability in the switching probability of the nanomagnet, with potential applications in probabilistic neuromorphic computing. 
\end{abstract}

\pacs{Valid PACS appear here}
\maketitle


Three-dimensional topological insulator surfaces enjoy gapless topological surface states (TSS) protected by spatial inversion and time-reversal symmetry. In many ways, they are similar to low-energy graphene bands, except the Dirac cones are labeled by spins rather than pseudospins, with a similar Chern number \cite{chen2009experimental,xia2009observation,zhang2009topological}. Numerous novel physical phenomena have been explored with TSS across graphene PN junctions. In particular, we can draw inspiration from graphene PN junctions that have served as a convenient laboratory for Klein tunnel devices, with potentially important contributions to low-power digital as well as high-speed analog switches  \cite{beenakker2008colloquium,stander2009evidence,sajjad2013manipulating, reza_modeling_2016,chiba_magnetic-proximity-induced_2017,fert_efficiency,SOT3dTI}. 

The intrinsic spin-momentum locking of TSS in a TI material allows conversion of a charge current into a spin current at a higher efficiency than heavy metal underlayers in common spin-orbit torque devices \cite{Fan2014,PRLSOT,SOTfert}. In fact, the spin-momentum locking feature of TSS imposes spin conservation rules across a PN junction, suggesting a  charge-to-spin conversion ratio (scaled dimensionally like the Spin Hall Angle except between longitudinal current densities) can exceed unity and be as high as 20 \cite{habib2015chiral}. This number is much higher than the Spin Hall Angle measured in heavy metals such as Pt ($0.07$)\cite{liu2011spin}, $\beta$-Ta ($0.12-0.15$)\cite{liu2012spin}, and in Pt-doped Au ($0.12$)\cite{gu2010surface}. However, besides an idealized assumption about the presence of a large non-leaky bandgap (perhaps more suitable to TIs like SmB$_6$ than Bi$_2$Se$_3$), it is also worth emphasizing that the spin `amplification' effect was studied on a homogeneous TIPNJ, where the source/drain contacts are assumed to be extensions of the TI surface as well. While it is easy to over-generalize the high charge-to-spin conversion efficiency to the TI-ferromagnet interface, it is not a straightforward equivalence to the homogeneous TI surface and deserves proper analysis. 

\begin{figure}[ht]
\includegraphics[width=3.25in]{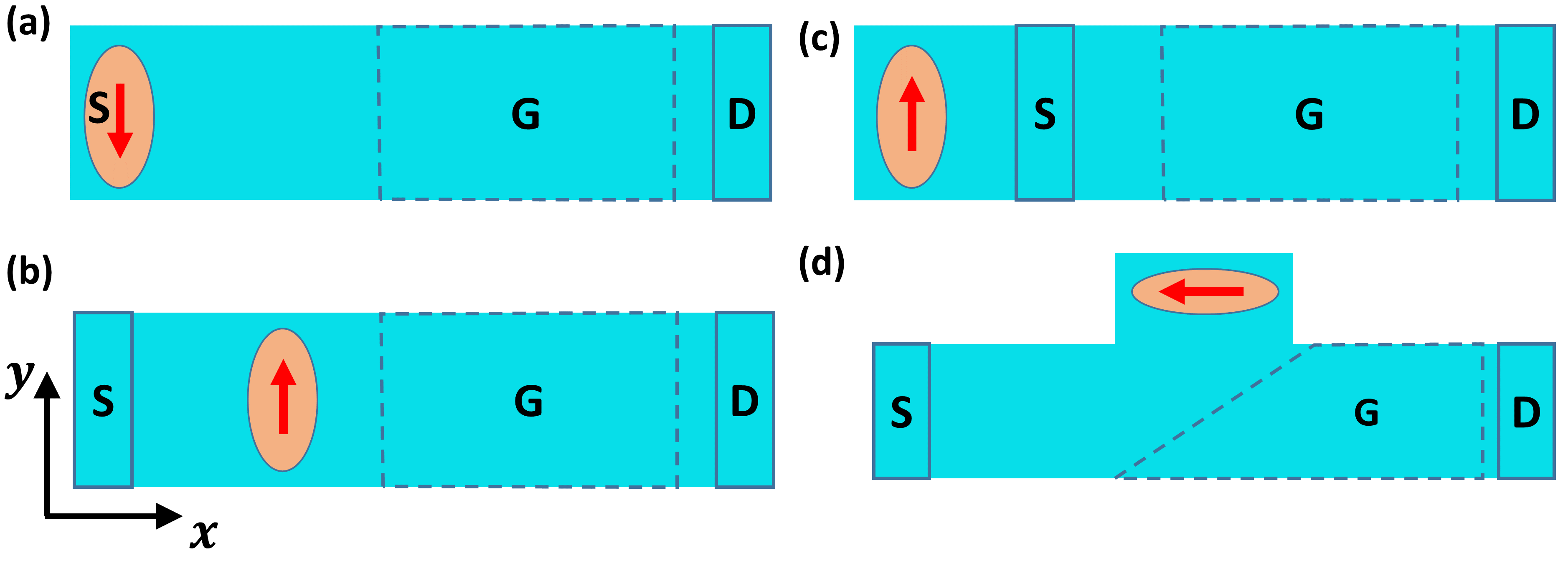}
\caption{Top view of different configurations of nanomagnet on a TIPNJ. (a) Nanomagnet as the source contact. (b) Floating nanomagnet located between source and drain contacts. (c) Floating nanomagnet outside the source-drain path. (d) A variation of the out-of-location floating nanomagnet. The gate (PN junction interface) is at $45^\circ$ degree from the source-drain direction. The angled PN junction can reflect electrons from the source to the nanomagnet.}
\label{fig:TIPNJ_magnet_config}
\end{figure}

\begin{figure}[th]
    \centering
    \includegraphics[width=3.5in]{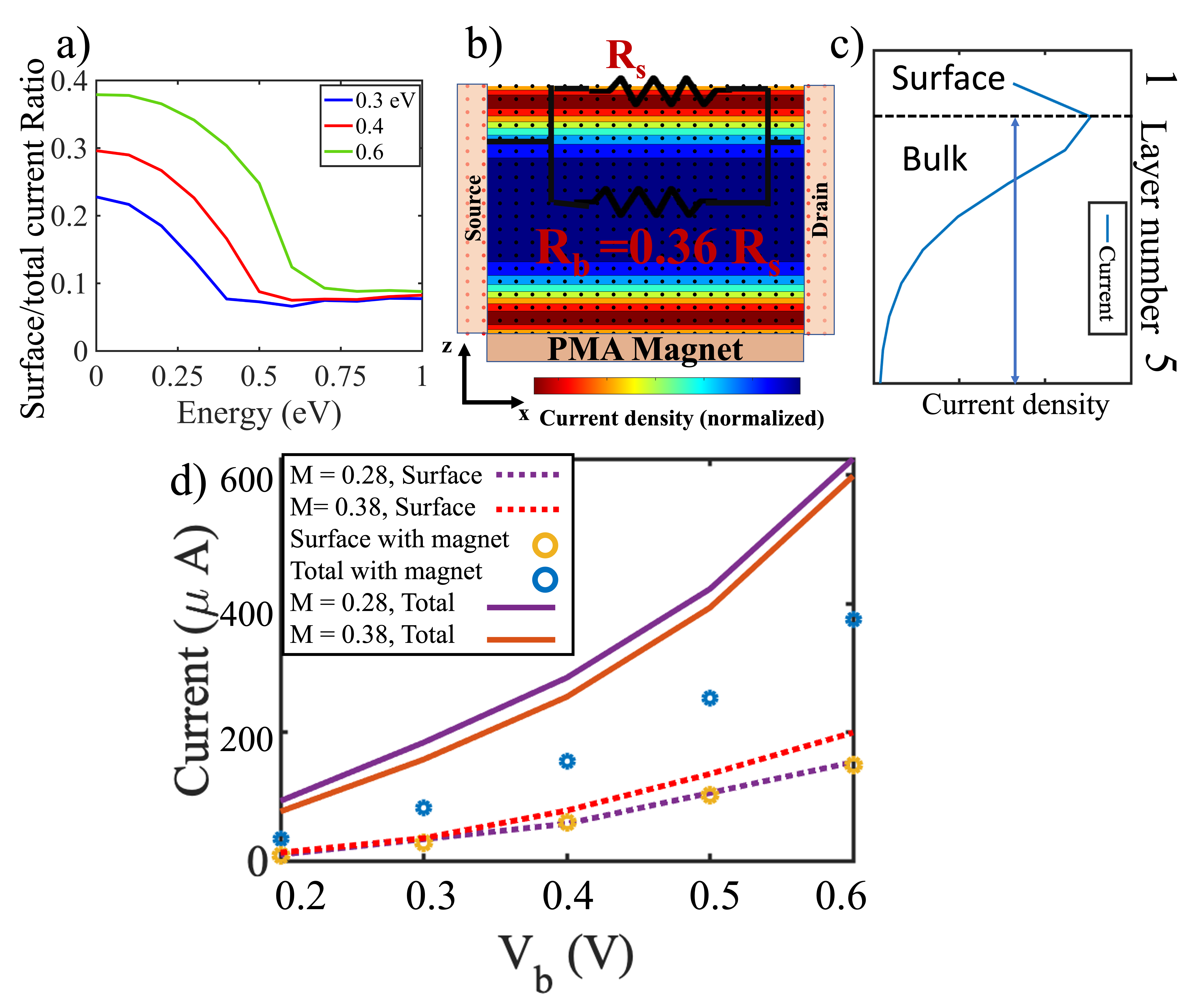}
    \caption{a) The ratio of TSS-to-Bulk current contributions as a function of the bulk band gap for various carrier energies. We simulate different bulk bandgaps by varying M 0.28, 0.38 and 0.55 $eV$ . For electron energies inside the bulk bandgap, the TSS dominate. We see that for energies above the band gap the TSS is weakly localized. b) Current distribution and a corresponding simple resistance model for $\mathrm{Bi_3Se_2}$ (with a bulk gap of 0.3 eV) at 0.2 V applied bias. The magnetic layer with perpendicular anisotropy (PMA) inserted between the bottom TI surface and the substrate (not shown) shows a possible way to lower the shunting effect. c) The surface current is defined as the current in the top layers; from the top layer to the layer with peaked current density. d) Two ways to improve the surface-to-bulk current ratio: with a larger bulk band gap, the surface current increases while the total current decreases. Alternatively, by inserting a magnetic layer with perpendicular anisotropy at the bottom surface, the total current is reduced while the current at the opposite surface is almost unaffected, translating to a higher contribution for the top surface. $V_s$ and $V_d$ are $\pm V_b/2$}.
    \label{fig:shunt}
\end{figure}

In this paper, we argue that the aforementioned large spin amplification is an intrinsic effect that does not automatically induce a large spin current across the TI/nanomagnetic contact interface. Fig.~\ref{fig:TIPNJ_magnet_config} shows some possible setups for a nanomagnet on TIPNJ. Our focus is on configuration (a) and we will use Non-Equilibrium Green's Function (NEGF) simulations on the TSS to help clarify the difference between intrinsic versus extrinsic charge-to-spin gain in TIPNJ. We also show a NEGF treatment of a full 3D slab that allows us to compare the surface versus bulk shunt resistances. In practice, it can be a challenge to isolate the surface contribution from the bulk in common binary TI compounds such as $\mathrm{Bi_2Se_3}$, $\mathrm{Bi_2Te_3}$\cite{taskin2009quantum, butch2010strong}. One possible solution is to alloy the binary TIs into ternary compounds like $\mathrm{Bi_2Te_2Se}$ tuned to have a low carrier density in the bulk \cite{jia2011low}. Another way to improve the TSS portion of the total current is by inserting a ferromagnetic layer with a perpendicular anisotropy to one surface (Fig \ref{fig:shunt}.b,c), which we will cover later. 

A PN junction provides a gate tunable filter on the spins. For sufficient current densities it can flip a low barrier in-plane magnet unidirectionally. The latter can be flipped back with an oppositely directed current in a symmetric set-up. The low barrier magnet is useful for three-terminal stochastic computing \cite{ASN,BSN}, where the PN junction provides a gate tunability of the average magnetization that follows a neuron-like nonlinear activation function. 

\begin{figure}
\includegraphics[width=8cm]{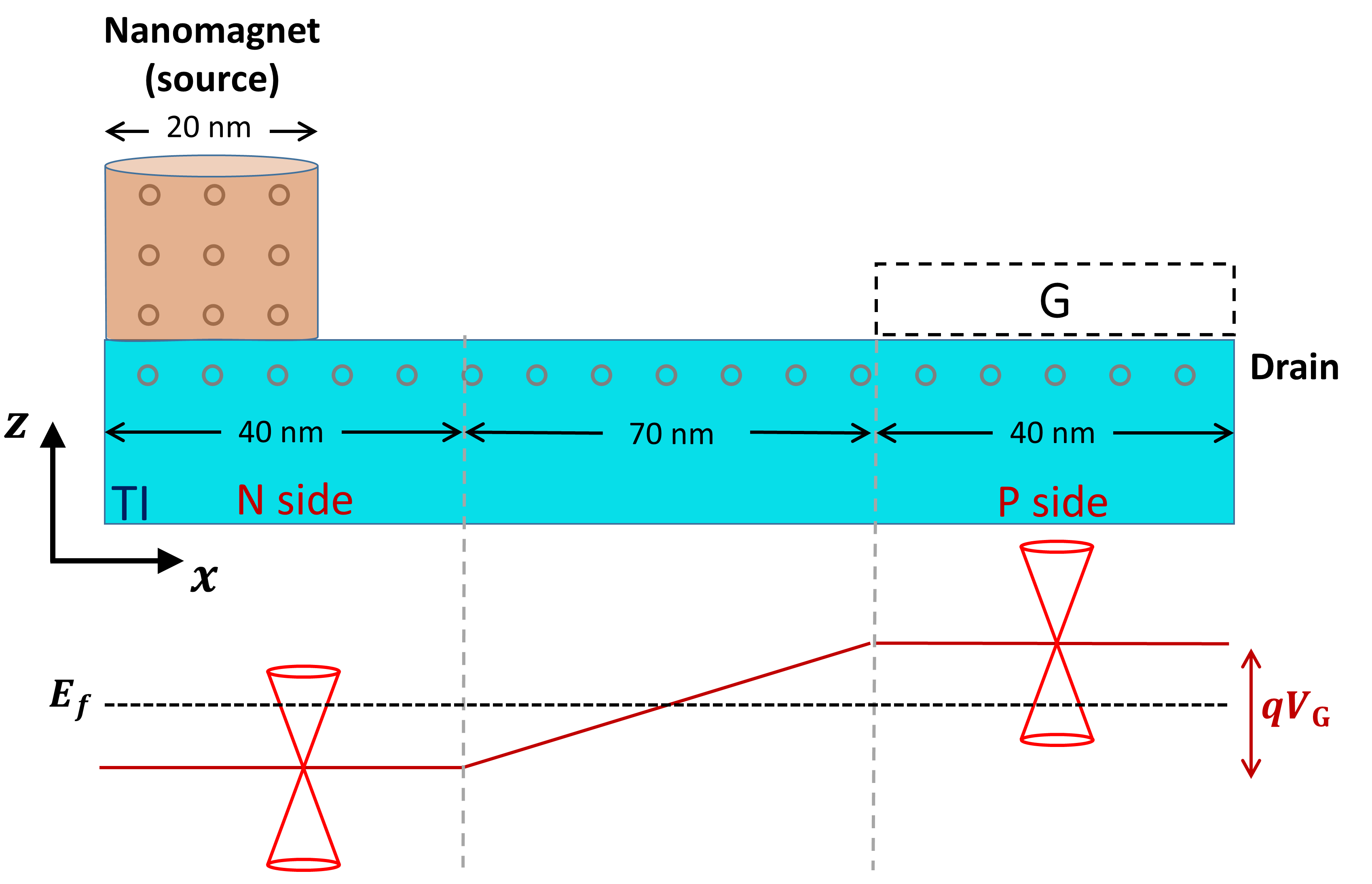}
\caption{Schematics of the discretization of a TIPNJ with a nanomagnet as the source contact. The ferromagnetic contact has dimension $20\,\mathrm{nm}$ along the x-direction. The TI surface is $150\,\mathrm{nm}$ long with $40\,\mathrm{nm}$ on the P and N side and $70\,\mathrm{nm}$ transitioning from N to P side. The gate on the drain side can swing the local TI surface to P-type. The red line represents the electrostatic potential of the TIPNJ.}
\label{fig:NEGF_grid} 
\end{figure}

\textbf{Computational Method.} Near the Dirac point, the 3DTI Hamiltonian can be described as \cite{zhang2009topological}:
\begin{equation}
    \epsilon_0(k)+\begin{pmatrix}
    \mathcal{M}(k) & A_1 k_z & 0 & A_2 k_-\\
    A_1 k_z & -\mathcal{M}(k) & A_2 k_- & 0\\
    0 & A_2 k_+ & \mathcal{M}(k) & -A_1 k_z \\ 
    A_2 k_+ & 0 & -A_1 k_z & -\mathcal{M}(k)
    \end{pmatrix}
\end{equation}
with $k_\pm={k_x\pm ik_y}$, $\epsilon_0(k)=C+D_1k_z^2+D_2(k_x^2+k_y^2)$ and $\mathcal{M}(k) = M- B_1k_z^2-B_2(k_x^2+k_y^2)$. 
For $\mathrm{Bi_2Se_3}$, the parameters used are: $M = 0.28~ \mathrm{eV}, A_1 = 2.3~ \mathrm{eV \mathring{A}}, A_2 = 3.3~ \mathrm{eV \mathring{A}}, B_1 = 6.8~ \mathrm{eV \mathring{A}^2}, B_2 = 44.5~ \mathrm{eV \mathring{A}^2}, C = -0.0068 ~\mathrm{eV}, D_1 = 5.74~ \mathrm{eV \mathring{A}^2}, D_2 = 30~ \mathrm{eV \mathring{A}^2}$. These parameters yield a bulk bandgap of $\approx300 ~\mathrm{meV}$. We used the 3D Hamiltonian to study the surface versus bulk current. By discretizing the 3DTI Hamiltonian we get the lattice Hamiltonian:
\begin{eqnarray}
    H_\mathrm{3DTI} = \sum_i c^\dagger_i\varepsilon_\mathrm{3DTI}c_i + \sum_i\bigg(c^\dagger_{i,i,i}\frac{T_x}{2}c_{i+1,i,i}+ \nonumber \\ c^\dagger_{i,i,i}\frac{T_y}{2}c_{i,i+1,i} + c^\dagger_{i,i,i}\frac{T_z}{2}c_{i,i,i+1}+\mathrm{H.C.} \bigg)
    \label{3DTI}
\end{eqnarray}
Where $\varepsilon_\mathrm{3DTI} = (C- 2D_1/a^2-4D_2/a^2)I_{4\times4}+(M-2B_1/a^2-4B_2/a^2) I_{2\times2}\otimes \tau_z$, $T_{x,y} =B_2/a^2\;I_{2\times2}\otimes\tau_z + D_2/a^2\;I_{4\times4}-iA_2/2a\;\sigma_{x,y}\otimes\tau_x$, $T_z =B_1/a^2\;I_{2\times2}\otimes\tau_z + D_1/a^2\;I_{4\times4}-iA_1/2a\;\sigma_z\otimes\tau_x$, and H.C. is the Hermitian conjugate. The z-direction is set perpendicular to the quantum layers. While the 3DTI Hamiltonian was used to study the bulk-surface current distributions and current shunting, we need to adopt a simpler TSS model for simulating the charge/spin transport in the combined system of a TIPNJ and a nanomagnet (Fig.~\ref{fig:NEGF_grid}): $H_\mathrm{TSS}=v_F\mathbf{\hat{z}}\cdot\left(\boldsymbol{\sigma}\times \mathbf{p}\right)$. To discretize this Hamiltonian, a Wilson mass term needs to be added to avoid the fermion doubling problem \cite{hong2012modeling}:
\begin{equation}
\begin{split}
H_\mathrm{TSS}&=v_F\mathbf{\hat{z}}\cdot\left(\boldsymbol{\sigma}\times \mathbf{p}\right) + \gamma\hbar v_F\sigma_z(k_x^2+k_y^2) \\
&=\sum_i c^\dagger_i\varepsilon_\mathrm{TI} c_i + \sum_i \left( c_{i,i}^\dagger t_x c_{i,i+1} + \mathrm{H.C.} \right) \\
&\qquad\qquad\quad + \sum_j\left(c^\dagger_{j,j}t_yc_{j,j+1}+\mathrm{H.C.}\right)
\end{split}
\label{eq:TI_H}
\end{equation}
where $\varepsilon_\mathrm{TI}=-4\hbar v_F{\alpha}\sigma^z/a$, $t_x=\hbar v_F[{i}\sigma^y/2+{\alpha}\sigma^z]/a$, $t_y=\hbar v_F[-{i}\sigma^x/2+{\alpha}\sigma^z]/a$. $a=5\,\mathrm{\AA}$ is the grid spacing, $v_F=0.5\times10^6\,\mathrm{m/s}$, and $\alpha=\gamma/a$ is a fitting parameter set to $\alpha=1$ to generate a bandstructure that reproduces the ideal linear bands within a energy window of $0.5 \,\mathrm{eV}$ \cite{habib2015chiral}. $\mathbf{\hat{z}}$ is the normal vector to the surface and $v_F$ is the speed of electrons near the Dirac point. $\boldsymbol{\sigma}=(\sigma_x,\sigma_y,\sigma_z)$ are the Pauli matrices.

A generic ferromagnetic (FM) nanomagnet is modeled by a tight-binding Hamiltonian in a cubic lattice with a single orbital per site:

\begin{equation}
\begin{split}
H_\mathrm{FM} =& \sum_{i,\sigma,\sigma'}c^\dagger_{i\sigma}\left(\varepsilon_\mathrm{FM}\delta_{\sigma\sigma'} -\frac{\boldsymbol{\Delta}}{2}\cdot\boldsymbol{\sigma} \right)c_{i\sigma'}\\
&\qquad -t\sum_{i,i',\sigma}\left(c_{i\sigma}^\dagger c_{i'\sigma}+\mathrm{H.C.}\right)
\end{split}
\end{equation}
where $\varepsilon_\mathrm{FM}$ is the onsite energy. $\boldsymbol{\Delta}=\Delta\mathbf{m}$ is the exchange energy split along the direction of the magnetization $\mathbf{m}$. $i'$ goes through all neighbors of site $i$. $t=\hbar^2/{2m^*a^2}$ is the electron hopping energy with effective electron mass $m^*$. The hopping term between the FM and TI surface is tuned to $t_\mathrm{FM-TI}=2.3t$ to minimize the contact resistance. The FM parameters $m^*=0.5m_e,\,\varepsilon_\mathrm{FM}=1.3\,\mathrm{eV},\Delta=0.8\,\mathrm{eV}$ result in a spin polarization $\eta=(D_\uparrow-D_\downarrow)/(D_\uparrow+D_\downarrow)\approx 0.57$ around the Fermi energy with density of states $D_\uparrow=1.34\times10^{46}\,\mathrm{J^{-1}m^{-3}}, D_\downarrow=0.357\times10^{46}\,\mathrm{J^{-1}m^{-3}}$ for the spin up ($-y$)/down ($+y$) channels. These numbers are a bit lower than the nanomagnet modeled in \cite{roy2015magnetization}, mostly due to the reduced size of our simulated magnet. Fig.~\ref{fig:NEGF_grid} shows the discretization of the coupled system including the ferromagnetic contact and the TI surface. The TI surface is assumed to be doped N-type with a single gate controlling the drain side. The periodic boundary condition is adopted in the $y$ direction and characterized by the transverse quasi-momentum $\mathbf{k}_\perp$.

 Part of the FM contact is included in the channel Hamiltonian while the rest is assumed to be infinite in the +z direction and modeled by a self-energy $\Sigma_S$ using recursive Green's functions. For simplicity, the drain is just an extended part of the TI surface and included as a surface self-energy term $\Sigma_D$ that can be obtained with an iterative approach \cite{sancho1985highly}. The retarded Green's function $G^r$ and electron correlation matrix $G^n$ can be calculated through the Non-Equilibrium Green's function method  \cite{datta1997electronic}
\begin{eqnarray} 
G^r(E,\mathbf{k_\perp})&=&[EI - H(\mathbf{k_\perp})-\Sigma_S(E,\mathbf{k_\perp})-\Sigma_D(E,\mathbf{k_\perp})]^{-1} \nonumber \\ 
G^n(E,\mathbf{k_\perp})&=& G^r\left(f_S\Gamma_S+f_D\Gamma_D\right)G^{r\dagger} \\
\Gamma_{S,D}&=&i\left(\Sigma_{S,D}-\Sigma_{S,D}^\dagger\right)
\end{eqnarray} 
with $f_{S,D}$ the Fermi-Dirac distribution function on the source/drain sides, and $I$ the identity matrix. The charge current is conserved throughout the structure and can be evaluated as: 
\begin{equation} 
I_q(E) = \frac{q}{h} \sum_\mathbf{k_\perp}\mathrm{Tr}\left[\Gamma_S G^r \Gamma_D G^{r\dagger}\right](f_S-f_D) 
\end{equation} 
while the charge and spin current from site $i$ to $j$ are calculated as:
\begin{eqnarray}
    \mathbf{J}^{i\rightarrow j}_s(E) &=& \frac{q}{ih} \sum_\mathbf{k_\perp}\mathrm{Tr}\left[\boldsymbol{\sigma}\left(H_{ij}G^n_{ji}-G^n_{ji}H_{ij}\right]\right)\nonumber\\
    \mathbf{J}^{i\rightarrow j}_q(E) &=& \frac{q}{ih} \sum_\mathbf{k_\perp}\mathrm{Tr}\left[\left(H_{ij}G^n_{ji}-G^n_{ji}H_{ij}\right]\right)
    \label{currentdensity}
\end{eqnarray}
The bias induced carrier density $n = n_\mathrm{neq} - n_\mathrm{eq}$ where $n_\mathrm{neq},n_\mathrm{eq}$ are obtained from:
\begin{eqnarray}
    n_\mathrm{eq} &=& -\frac{1}{\pi}\mathbf{Tr}\left(\sum_\mathbf{k_\perp}\int dE \mathbf{Im}(G^r (E,k_\perp)) \right)f_0(E)\nonumber \\
    n_\mathrm{neq} &=& \frac{1}{2\pi i}\mathbf{Tr}\left(\sum_\mathbf{k_\perp}\int dE G^n (E,k_\perp) \right)
\end{eqnarray}
where $f_0$ is the equilibrium Fermi-Dirac distribution.

\begin{figure}[th]
\includegraphics[width=3.4in]{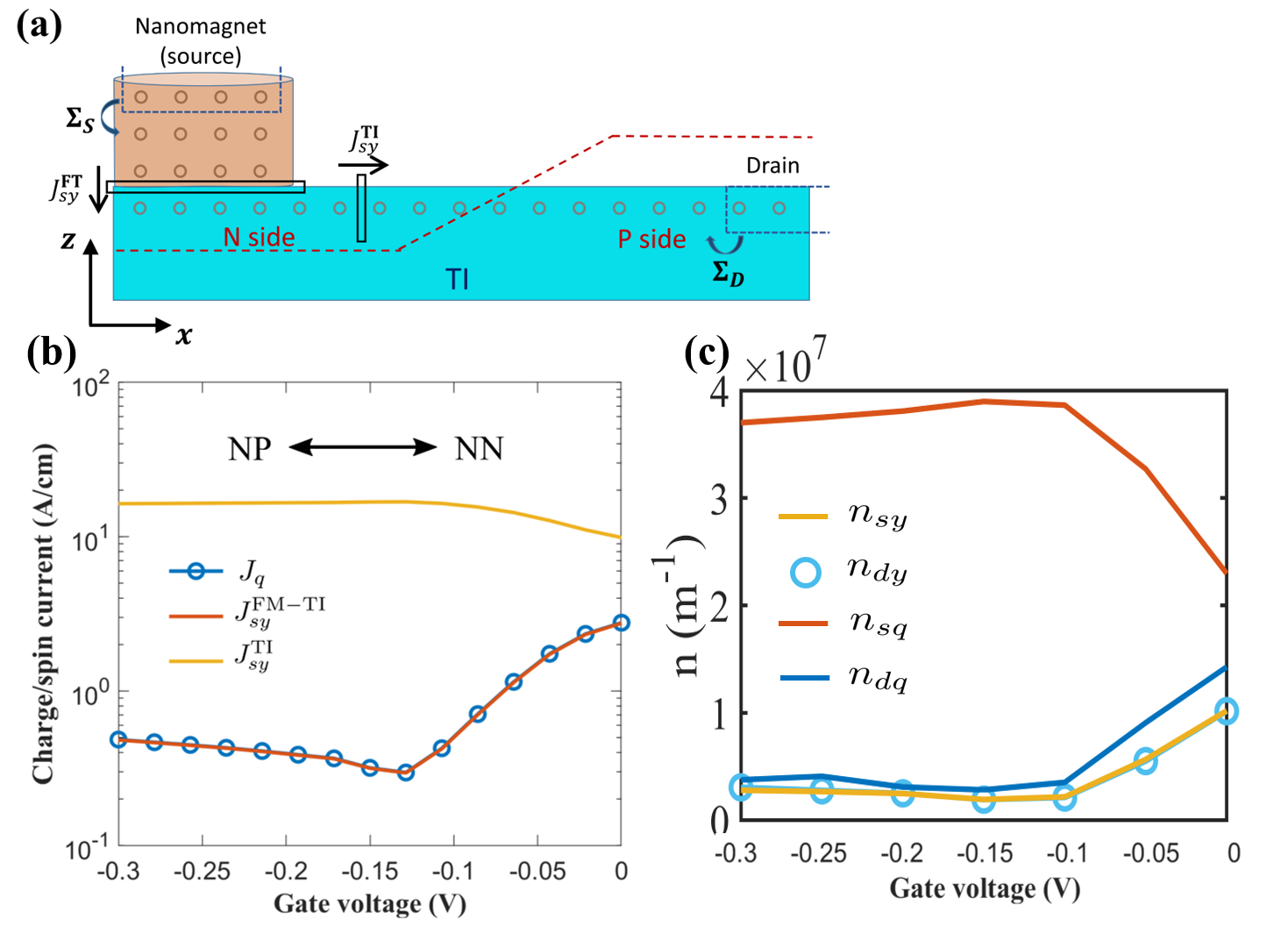}
\caption{I-V characteristic of TIPNJ with a ferromagnetic source contact. (a) TIPNJ setup with a nanomagnet as the source. The magnetization of the FM contact is oriented to the $-y$ direction. The gate contact is present but not visualized in the schematic. (b) Charge and spin current densities calculated at different locations. $J_q$ is the charge current density (conserved throughout the system). $J_{sy}^\mathrm{TI}$ is the spin current (polarized along $-y$ direction) calculated on the TI surface on the source side (N side). $J_{sy}^\mathrm{FM-TI}$ is the spin current density at the FM-TI interface. The source-drain bias $V_{sd}$ is fixed at $0.1\,\mathrm{V}$ and the magnetization of the nanomagnet is aligned in the $-y$ direction. (c) Spin density shows almost no change with gate voltage and is the same in the source and drain side of TI. This indicates that the spin current amplification in the source region of TI can not be used for magnet switching.}
\label{fig:FM_TI_IV}
\end{figure}

{\bf{Results: charge-to-spin conversion in TIPNJ.}}
Our previous work\cite{habib2015chiral} showed that the TIPNJ acts like a spin collimator that increases the non-equilibrium spin current while reducing the charge current at the same time. The corresponding spin-to-charge ratio (the longitudinal equivalent of the Spin Hall Angle $\theta_\mathrm{SH}$) at the source contact can go up to as high as $20$ \cite{habib2015chiral}. While it is easy to assume that this conversion would dramatically improve the switching efficiency if we replace the `TI source contact' with a ferromagnetic contact, we show here that the impressive gain is limited to the TI surface (we refer to this as the `intrinsic' gain).
To see the difference between the intrinsic and the external gain, we calculated the spin current at two locations along the transport direction: one between the FM contact and the TI surface, the other one on the N-type TI surface, as indicated in Fig.~\ref{fig:FM_TI_IV}(a). Fig.~\ref{fig:FM_TI_IV}(b) compares the charge current with the spin currents at the above-mentioned locations as a function of the gate voltage. As the gate voltage sweeps from $-0.3\,\mathrm{V}$ to $0\,\mathrm{V}$, the TI surface transitions from a PN junction to a homogeneous N-type surface. The behavior of the charge current (independent of where it is being calculated) and TI surface spin current $J_{sy}^\mathrm{TI}$, with a maximum ratio of $J_{sy}^\mathrm{TI}/J_q > 20$ resembles the results from \cite{habib2015chiral}, where the PN junction effectively suppresses the charge current while amplifying the {nonequilibrium} spin current. However, the spin current across the nanomagnetic contact $J_{sy}^\mathrm{FM-TI}$ does not follow the in-plane surface spin current but instead follows the charge current. While the ratio $J_{sy}^\mathrm{FM-TI}/J_q$ is very close to $100\%$, $J_{sy}^\mathrm{FM-TI}$ never exceeds the charge current regardless of which regime the TI surface is in. Looking at the bias-induced charge and spin carrier density, we can see the picture more clearly (Fig.~\ref{fig:FM_TI_IV}.c). The spin density shows no change on the source and drain sides of TI. However, the charge densities show significant variations in the source and drain regions (opposite of currents' behavior). This further solidifies the above explanation, as the charge densities from the incoming and reflected currents add up in the source region but are subtracted for spin density.
\begin{figure}[ht]
\includegraphics[width=3.5in]{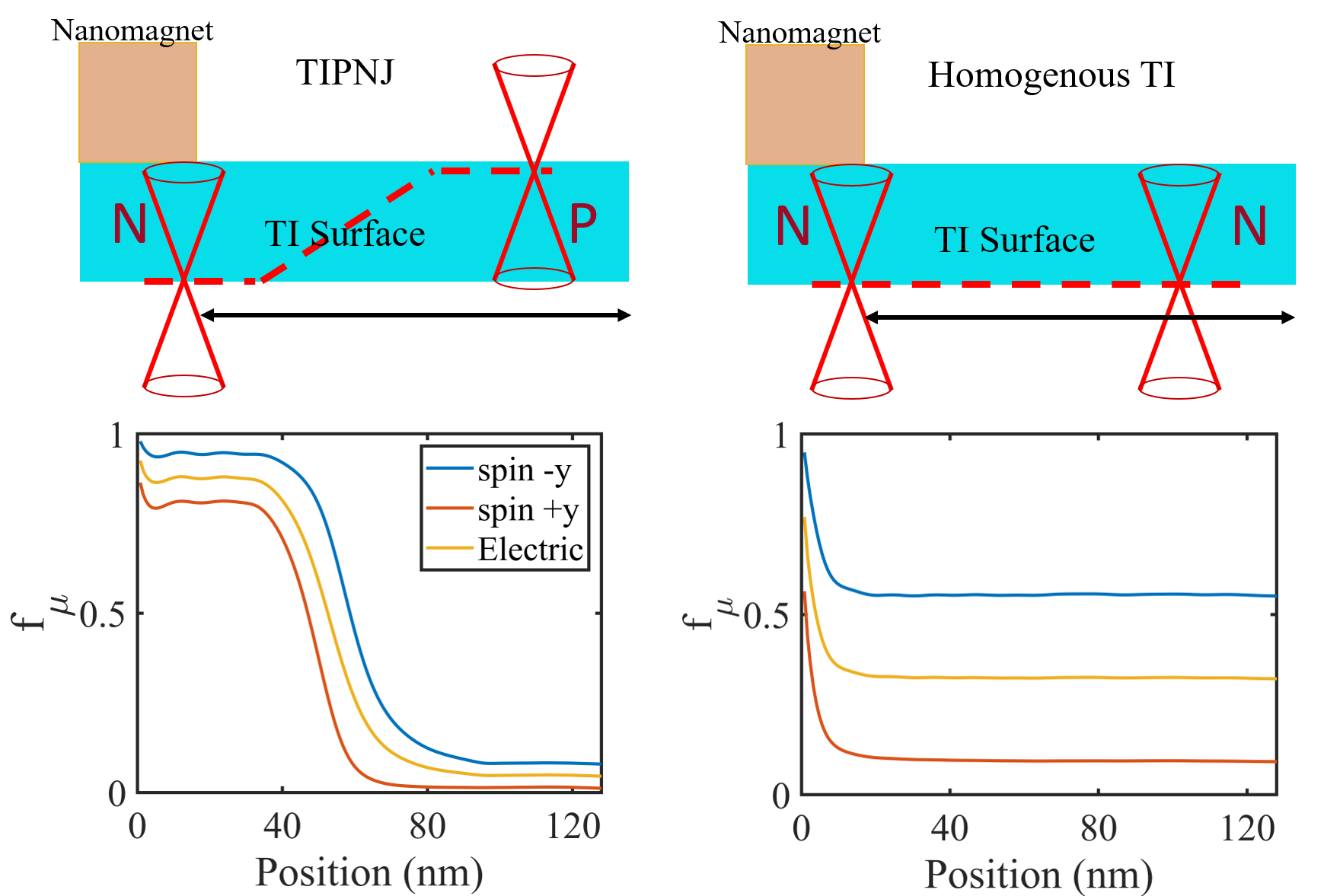}
\caption{Fermi levels $f_\mu$ for different spin channels on the TI surface in different regimes (TIPNJ versus homogeneous TI surface). $\mu_s (\mu_d)$ is the Fermi level of the source (drain) contact. As can be seen from the comparison, PN junction raises the Fermi level of all spin channels on the source side due to the reflection of the junction potential barrier. The biggest potential drop happens near the junction interface. In the case of homogeneous TI, the Fermi level remains uniform throughout the TI surface from the source to the drain.}
\label{fig:TI_fermi_spacial}
\end{figure}
To better understand this discrepancy, we look at the individual Fermi levels $f_\mu$ in each spin channel (where spin-up corresponds to $-y$ while spin-down corresponds to $+y$) on the TI surface (Fig.~\ref{fig:TI_fermi_spacial}). The Fermi levels are calculated by simulating a magnetic probe at each lattice point which draws net zero currents from the channel with negligible deformation of the channel Hamiltonian. The calculated quantity is {$f_\mu = \frac{Tr (\Gamma_\mathrm{FM} G^n)}{Tr(\Gamma_\mathrm{FM} A)}$} \cite{hong2012modeling} where $\Gamma_\mathrm{FM}(\pm y) = \gamma_m(I \pm \sigma_y)$ is the coupling between FM probe and the TI surface with $\gamma_m \ll \hbar v_f/a$ being the coupling strength. Due to the spin-momentum locking of the TI surface states, electrons in the spin-up channel can only move in the $+x$ direction while electrons in the spin-down channel move in the opposite direction, {roughly along} two one-way streets {(for 3D TSS, the angular transition is continuous)\cite{yunkunklein}}. This behavior results in the `intrinsic' spin amplification observed on the TI surface: non-equilibrium electrons moving in opposite directions due to electron reflection at the PN junction reduce the net charge current while increasing the net {nonequilibrium} spin current. However, when the nanomagnet contact exchanges electrons with the TI surface, the electrons flow bidirectionally regardless of their spin orientations. Ultimately, the rate of influx and outflow of spins from the nanomagnet is controlled by the difference in the electrochemical potentials between the magnet and the TI surface. Being the source contact, the magnet has the highest electrochemical potential in both spin channels. Therefore one expects a net outflow of electrons from the magnet to the TI surface in both spin channels.  The net spin current is just the difference between the charge currents in different spin channels, which is always lower than the sum of them. The extrinsic spin amplification {in this configuration, with the magnet doubling as the current source,} cannot exceed unity.



The limit on external charge-to-spin gain is imposed by having the nanomaget as the source contact. A more common setup is to have a floating nanomagnet on top of a TI (see fig.~\ref{fig:TIPNJ_magnet_config}(b)). There are several research papers that discuss this configuration (without a PN junction) in detail, such as the proximity effect induced by the magnetic field from the magnet \cite{duan2015nonlinear}, and the shunting of the charge current from the TI surface to a conductive magnetic layer \cite{roy2015magnetization}. With this configuration, the electrochemical potential of the nanomagnet is free to adjust in order to draw a net zero charge current.

The charge-to-spin conversion rate depends on the splitting of the electrochemical potentials of the opposite spin channels. With TIPNJ, Fig.~\ref{fig:TI_fermi_spacial} shows that the splitting narrows compared to the homogeneous TI surface, which means the PN junction effectively turns off the spin torque from the TI surface.

Because of the current shunting to the bulk of the 3DTI, the surface spin current calculated from TSS Hamiltonian Eq.~\ref{eq:TI_H} is overestimated. To study the surface-bulk current distribution, the 3DTI Hamiltonian as described in Eq.~\ref{3DTI} is employed. With equation \ref{currentdensity}, the current density for each layer is calculated and separated into bulk and surface currents. To get a simple resistance model, we use two parallel resistances for bulk $R_b$ and surface $R_s$. Based on the current distributions, the ratio of resistances is calculated. Here we define the surface current as the current in the top (bottom) layers: starting from the top (bottom) surface layer to the layer where the current density is peaked as shown in Fig.~\ref{fig:shunt}.c. TSS has a practical limitation on how much current increases with applied bias. For electrons with higher energies that are only accessible with a higher bias, the conductive state will no longer be strongly localized at the surface and will shunt into the bulk (Fig.~\ref{fig:shunt}(a)). Such a shunting \cite{3dTIacoustic} implies that a significant portion of the current will be diverted away from the surface, reducing the carrier exchange between the FM and TI and degrading the switching efficiency.
Increasing the bias above the bulk bandgap will therefore have diminishing returns. For example, for $\mathrm{Bi_2Se_3}$ the localized current at the surface (about $2\mathrm{nm}$ thick, Fig.~\ref{fig:shunt}.b) carries only $30 \%$ of the total current. 
One way to alleviate this is by finding a 3DTI material with a larger bulk bandgap, as can be seen in Fig~\ref{fig:shunt}(a-b). Alternatively, inserting or doping the TI surface in contact with the substrate (Fig.~\ref{fig:shunt}b,d) the 3DTI grown on with a PMA magnet can also increase $\theta_\mathrm{SH}$. This is due to the fact that out-of-plane Zeeman energy \cite{kane,nikolic3dti, chiba_magnetic-proximity-induced_2017, tokura_magnetic_2019} (Added as $ m_z \sigma_z\otimes I_{2\times2}$ to TI onsite energy for the bottom surface) opens up a gap. Since this effect is localized only at one surface, the opposite surface is nearly unaffected. However, the total current is now smaller as the total resistance of the channel is increased.

\begin{figure}[ht]
    \centering
    \includegraphics[width=0.8\columnwidth]{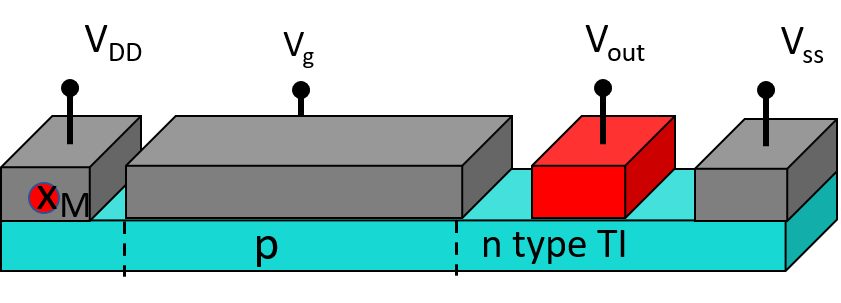}
    \caption{Figure shows the structure of a 1T-1MTJ memory based on TIPNJ. The TIPNJ acts as the transistor element of the memory. $V_g$ is the gate voltage which controls the type of TI (P or N), which then can tune the TSS accordingly. The $V_\mathrm{out}$ is the output voltage which depends on the magnetization direction.}
    \label{fig:TIPNJ}
\end{figure}

\textbf{TIPNJ as a RAM.}
For the nanomagnet-as-source-contact configuration sans external gain, the gate controls the TIPNJ resistance and thus can be used to tune the total spin current and the spin torque applied on the nanomagnet. The structure of such a device is shown in Fig.~\ref{fig:TIPNJ}, where the TIPNJ would have a transistor-like behavior. However, the maximum bias window is limited by the bandgap of the bulk TI. While we have discussed ways to alleviate the shunting issue, they might be challenging to achieve experimentally. Having said that, we looked at a possible 1 transistor-1 magnetic tunnel junction (1T1MTJ) \cite{1T1MTJ} device based on TIPNJ.
The switching of a magnet parallel or anti-parallel to the injected spin current polarization is known as the anti-damping switching. The required critical spin current is given by (ignoring the effects of the rotation of spins from interfacial Zeeman field) \cite{sun}:
\begin{equation}
    I_s \approx \frac{4q\alpha k_BT\Delta}{\hbar}
    \label{eq:Js}
\end{equation}

\noindent where $\Delta$ is the dimensionless energy barrier and $\alpha$ is the Gilbert damping. From NEGF calculations, we can see that at low biases, spin currents of $100~\mathrm{\mu A}$ at a width of 40 $\mathrm{nm}$ is achievable. This would be enough to switch a magnet with $\Delta = 40,\;$ (needed for a reliable memory at room temperature \cite{liu2012spin}), $\alpha = 0.1$ at a write error rate (WER) of $10^{-9}$ and a switching time of $5 \mathrm{ns}$. The switching roughly translates to an energy consumption of $\approx 600\;\mathrm{fj/bit}$. It has been reported that 3DTI can be more energy efficient than heavy metals \cite{PRLSOT,Liu_ti}. However, by improving the interfacial quality of FM and heavy metal, a significant increase in efficiency of heavy metal has also been reported \cite{fert_efficiency,improvedHM}. This suggests that although 3DTI might not provide higher energy efficiency compared to heavy metals, TIPNJ can operate at low voltages suitable for more energy efficient applications such as RAM. For instance, the tuning of the TI surface state can be achieved by small applied voltages ($V_d \leq 0.1 - 0.5 V$ as compared to $0.9 V$ for CMOS transistors).
From a practical perspective, the required on/off current ratio $\beta$ from reliable spin torque switching can be deducted from the error rate \cite{butler,fokker}:
\begin{equation}
    \beta = {\log\Bigl[\frac{\pi ^2 \Delta}{4P_e}\Bigr]}/{\log\Bigl[\frac{\pi ^2 \Delta}{4(1-P_e)}}\Bigr]
\end{equation}
For a typical write error rate $P_e = 10^{-9}$, the minimum $\beta$ is about 5.5. In comparison, the $\beta \approx 8$ calculated from NEGF simulations shows that the gating mechanism works well enough for a TI based 1T-1MTJ RAM. 
\begin{figure}[th]
\includegraphics[width=\columnwidth]{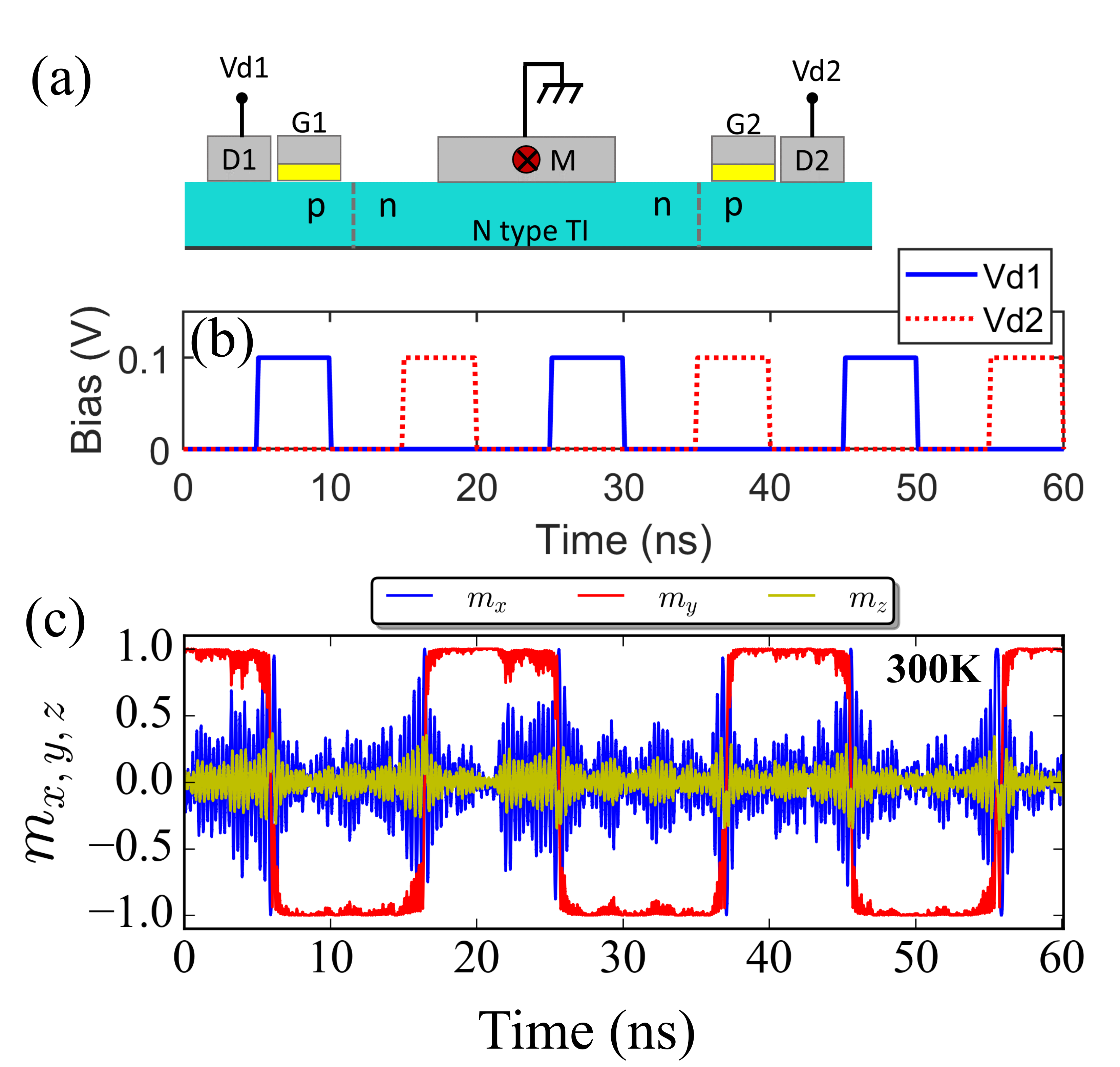}
\caption{ (a) Proposed experimental set-up to test gate tunable up to down (1 to 0) switching and vice-versa. Dual TIPNJ setup for switching the nanomagnet back and forth. (b) Voltage pulse profile. (c) Stochastic Landau Lifshitz Gilbert (LLG) simulation of magnetodynamics of an in-plane nanomagnet ($40\times 20\times 4\,\mathrm{nm}$) with $M_s=400\,\mathrm{emu/cc},\;\alpha=0.1$ with thermal stability $\Delta\approx5$ at 300K.  When the drain $V_{d1}$ is activated with a positive bias relative to the source, the spin torque flips the magnet's in-plane magnetization ($m_y: +1 \rightarrow -1$), at which point further spin injection is blocked by spin-momentum coupling and the magnetization stays pinned. Subsequently activating the other TIPNJ with $V_{d2}$ flips the magnet back to the $-y$ orientation.}
\label{fig:LLG}
\end{figure}

\begin{figure}[th]
\includegraphics[width=3.25in]{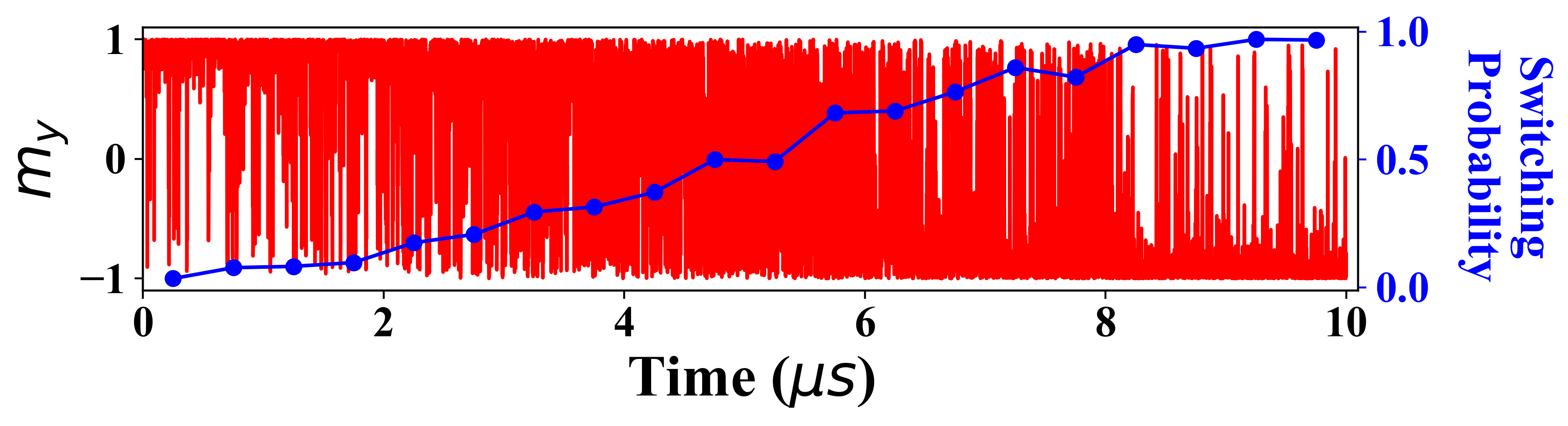}
\caption{Programming a magnet based binary stochastic neuron using gate tunable PN junction barrier. Low-energy barrier magnet operating in the stochastic regime. A nanomaget of dimension $20\,\mathrm{nm}\times 18\,\mathrm{nm}\times 5\,\mathrm{nm}$ with approximately $\sim 1k_BT$ energy barrier. The current is applied as a linear ramp from $+5\,\mathrm{\mu A}$ to $-5\,\mathrm{\mu A}$ where negative current means the current is flowing through the other PN junction, which can be achieved by tuning the drain or gate bias.}
\label{fig:sto_LLG}
\end{figure}

\textbf{Volatile Memory.} Although it is possible to enhance the TIPNJ current through material engineering to achieve the required current to drive traditional RAM applications, we also explored what is possible under conditions where leakage current and interfacial imperfections lower the spin current generated on a TIPNJ. To quantify the driveability of the TIPNJ, we have simulated the magnetodynamics of a nanomagnet under the TIPNJ current with the phenomenological Landau-Lifshitz-Gilbert (LLG) equation {with a stochastic Langevin correction}:
\begin{equation}
\begin{split}
 \frac{d \mathbf{m}}{d t}=& -\frac{\mu_0\gamma}{1+\alpha^2}\mathbf{m}\times\mathbf{h}-\frac{\alpha\mu_0\gamma}{(1+\alpha^2)}\mathbf{m}\times\left(\mathbf{m}\times\mathbf{h}\right)\\ & -\frac{\mu_B}{q\Omega M_s(1+\alpha^2)}\mathbf{m}\times\left(\mathbf{m}\times\mathbf{I}_s\right)\\
 \mathbf{h} =& \mathbf{H}_\mathrm{eff}+\boldsymbol{\tau}_s,~~\boldsymbol{\tau}_s=\sqrt{\frac{2\alpha k_BT}{\mu_0\gamma\Omega M_s}}\mathbf{G}
\end{split}
\label{eq:LLGS}
\end{equation}
where $\mathbf{m}$ is the normalized magnetic moment; $\mathbf{H}_\mathrm{eff}$ is the effective field from the intrinsic anisotropy of the magnet; $M_s$ is the saturation magnetization; $\Omega$ is the volume of nanomagnet; $\mathbf{I}_s$ is the spin current with the vector pointing at the spin direction. $\boldsymbol{\tau}_s$ is a stochastic thermal torque term representing the thermal noise. $\mathbf{G}$ a three dimensional Gaussian white noise with mean $\langle G(t)\rangle=0$ and standard deviation of $\langle G^2(t)\rangle=1$. First, we propose a configuration (Fig.~\ref{fig:LLG}(a)) for an experimental verification of the switching of a nanomagnet from a TIPNJ. A symmetric structure is needed for flipping the magnetization back-and-forth, generating a periodic signal that can be read out by stacking an MTJ or spin valve structure on top of the switching magnet. Our simulation shows that the TIPNJ can reliably drive a small magnet with an adequate anisotropy energy barrier ($5k_BT$) to reliably switch back and forth with a current (proportional to the longitudinal dimension of the magnet) as low as $I_s=10\,\mathrm{\mu A}$. When we further lower the anisotropy barrier, we entered a regime where the tunability of the TIPNJ is favored over its driveability. A nanomagnet with very small energy barrier ($\sim 1k_BT$) can operate in the stochastic switching regime. At room temperature, those magnets flip their magnetic moments frequently and stochastically under the influence of thermal noise. A small current can tip the balance of the two energy minimum orientations along the easy axis and affect the probability of the magnetic moment pointing at the direction decided by the current direction, with a binary stochastic neuron like tanh behavior on average.  Fig.~\ref{fig:sto_LLG} shows the tuning of switching probability under a small spin current $I_s=5\,\mathrm{\mu A}$ generated from the TIPNJ. The state (probability) of the nanomagnet can be extracted through a time-average measurement on the magnetic moment. This kind of setup can be used in generating random numbers \cite{fukushima2014spin}, or stochastic signal processing \cite{ganguly2018analog}. The energy cost for this is calculated to be $\approx 20\;\mathrm{fj/bit}$.

{\bf{Conclusion.}}
In this paper, we have taken a closer look at the tuning of the TSS spin current in a TIPNJ as was originally proposed in \cite{habib2015chiral}. We have analyzed the intricate nature of the charge-to-spin conversion on the TI surface with a nanomagnetic source contact. More importantly, we have clarified an important difference between the intrinsic charge-to-spin gain observed on a homogeneous TI surface, and the external gain useful for switching a nanomagnet. Additionally, we have looked at the limitations on the magnitude of the surface current, which is mostly determined by the bulk bandgap, and discussed two ways to improve it. While those issues seem to suggest the TIPNJ might not be able to offer a superior solution for common memory/logic applications, it can still work well as a RAM because of its gate tunability and low bias operation. Furthermore, the spatial programmability introduced by the PN junction may find use in novel applications such as probabilistic or neuromorphic applications.

{\bf{Acknowledgments.}} This work is supported by the Army Research Lab (ARL) and in part by the NSF I/UCRC on Multi-functional Integrated System Technology (MIST) Center; IIP-1439644, IIP-1439680, IIP-1738752, IIP-1939009, IIP-1939050, and IIP-1939012. We acknowledge helpful discussions with Mahesh Neupane and George De Coster at ARL.
\bibliographystyle{unsrt}
\bibliography{ref}
\end{document}